\documentclass{iaus}
\usepackage{graphics,epsf,psfig,color}

  \checkfont{eurm10}
  \iffontfound
    \IfFileExists{upmath.sty}
      {\typeout{^^JFound AMS Euler Roman fonts on the system,
                   using the 'upmath' package.^^J}%
       \usepackage{upmath}}
      {\typeout{^^JFound AMS Euler Roman fonts on the system, but you
                   dont seem to have the}%
       \typeout{'upmath' package installed. iaus.cls can take advantage
                 of these fonts,^^Jif you use 'upmath' package.^^J}%
      }
  \else
  \fi

  \checkfont{msam10}
  \iffontfound
    \IfFileExists{amssymb.sty}
      {\typeout{^^JFound AMS Symbol fonts on the system, using the
                'amssymb' package.^^J}%
       \usepackage{amssymb}%

      }{}
  \fi

  \IfFileExists{amsbsy.sty}
    {\typeout{^^JFound the 'amsbsy' package on the system, using it.^^J}%
     \usepackage{amsbsy}}
    {\providecommand\boldsymbol[1]{\mbox{\boldmath $##1$}}}

\title[Outskirts of Galaxy Clusters: intense life in the suburbs]
      {Simulating the Formation of Galaxy Clusters}

\author[Daisuke Nagai and Andrey Kravtsov]%
{Daisuke Nagai and Andrey V. Kravtsov}

\affiliation{Department of Astronomy and Astrophysics, Kavli Institute for
  Cosmological Physics, The University of Chicago, Chicago, IL 60637;
  {\tt daisuke,andrey@oddjob.uchicago.edu}}

\pubyear{2004}
\volume{195}
\pagerange{1--8}
\date{?? and in revised form ??}
\jname{Outskirts of Galaxy Clusters: intense life in the suburbs}
\editors{A. Diaferio, ed.}
\begin{document}

\maketitle

\begin{abstract}
  We study the effects of radiative cooling, star formation and
  stellar feedback on the properties and evolution of galaxy clusters
  using high-resolution Adaptive Mesh Refinement N-body+gasdynamics
  simulations of clusters forming in the $\Lambda$CDM universe.
  Cooling leads to the condensation of gas in the inner regions of
  clusters, which in turn leads to steepening of the dark matter
  profile. The cooling gas is replaced by the higher-entropy gas from
  the outer regions, which raises the entropy and temperature of gas
  in the cluster core.  The magnitude of these effects is likely
  overestimated in the current simulations because they suffer from
  the overcooling problem: a much larger fraction of baryons is in the
  form of cold gas and stars than is observed.  We find that the
  thermal stellar feedback alone does not remedy this problem.
  Additional ad-hoc preheating can lower the amount of cold gas but a
  simple uniform preheating results in incorrect star formation
  history, as it delays the bulk of star formation until $z<1$.  Our
  analysis shows that the overcooling in a cluster as a whole is
  really the overcooling in the central galaxy and its progenitors at
  high redshifts.  This indicates that an additional heating mechanism
  that can continuously heat the gas in the cluster core is required
  to reproduce the observed cluster properties. Energy injection by
  the Active Galactic Nuclei, which may provide such heating, may thus
  be an important missing ingredient in the current theoretical models
  of cluster formation.
\end{abstract}

\firstsection 
\section{Effects of Cooling, Star formation and Feedback}
 
Below, we highlight some of the important effects of gas cooling and
star formation on the properties of galaxy clusters:
\vspace{2mm}

\begin{itemize}
  
\item Gas cooling induces condensation of gas in the cluster core
  region ($\lesssim 100$kpc), and the baryon mass in the cluster core
  at $z=0$ is dominated by the stars associated with the central
  galaxy.
  
\item The condensation of baryons causes contraction of dark matter
  and steepens the dark matter profile in the central region from
  $r^{-1.5}$ in the adiabatic runs to $r^{-2.0}$ in the simulations
  with cooling.
  
\item Gas cooling induces inflow of high-entropy gas from the outside and
  raises the level of entropy and gas temperature in the cluster core
  region.
  
\item Stellar feedback and/or additional preheating suppress the
  effects described above by regulating gas cooling. The magnitude of
  the effects on the distribution of dark matter, entropy and
  temperature of gas is, therefore, sensitive to the total fraction of
  baryons which cooled to form stars and cold gas clouds.

\end{itemize}

\begin{figure}
\centerline{ \epsfysize=2.7in \epsfbox{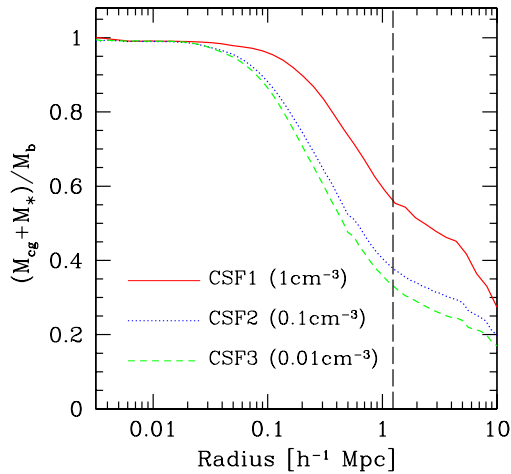} }
\vspace{-0.2cm}
\caption{
  The cold gas + stellar mass fraction of the simulated cluster that
  includes cooling, star formation and stellar feedback.  Three lines
  indicate the CSF runs with varying star formation : the star
  formation is allowed to proceed only in the region where the local
  gas density is larger than 1.0 (Red), 0.1 (Blue) and 0.01cm$^{-3}$
  (Green). The figure shows that 35-55\% of baryons within the virial
  radius (indicated by the dashed line) are in the form of cold gas
  and stars at $z=0$, which is significantly larger than the observed
  value of 10-20\%. }
\label{fcold}
\end{figure}
\section{The Overcooling Problem }

The cluster simulations with cooling and starformation revealed a
significant problem: 35-55\% of baryons within the virial radius are
in the form of cold gas and stars at $z=0$, which is significantly
larger than the observed value of $\lesssim$10-20\% ({\it e.g.,}
Balogh 2001; Lin, Mohr \& Stanford 2003). This problem was previously
reported in the simulations using the Smoothed Particle Hydrodynamics
(SPH) simulations ({\it e.g.,} Suginohara \& Ostriker 1998; Lewis
et.al. 2000; Tornatore et.al. 2003).  We find that the same problem
exists in our higher-resolution Eulerian simulations that include
metallicity-dependent radiative cooling.  Moreover, we find that the
cold and stellar mass fractions depend on the details of star
formation implementation.  However, we were not able to solve the
overcooling problem by varying parameters of thermal stellar feedback
within reasonable limits, as shown in Figure~\ref{fcold}.

\section{Effect of Preheating} 

Additional preheating can help resolve some of the discrepancies
between cluster simulations and observations.  Such preheating,
presumably associated with the epoch of vigorous starformation, is
often advocated as the solution to the overcooling problem and the
discrepancies between the predicted and observed cluster scaling
relations ({\it e.g.,} Bialek et.al. 2001; Muanwong et.al. 2002, da
Silva et.al. 2004). Our results, however, indicate that the additional
preheating results in starformation history significantly different
from observations, as shown in Figure~2.  If the preheating occurs
relatively early ($z\sim 3$) the gas manages to cool back by z$\sim$1,
and the resulting star formation rate remains high
($\sim 1000\rm M_{\odot}yr^{-1}$) at $z\lesssim 0.5$. This results in
incorrect star formation history: most of the cluster stars form in
the last 7-10 billion years in stark contrast to observations, which
indicate that less than 30\% of cluster stars form during this epoch.  If
preheating occurs late, the amount of the required energy is
unrealistic and the expected effect on the intergalactic medium is in
contradiction to observations of the low-redshift lyman alpha forest.
These results provide a strong argument against the simple uniform
``preheating'' scenario.

\begin{figure}
\centerline{ 
   \epsfysize=2.5truein  \epsffile{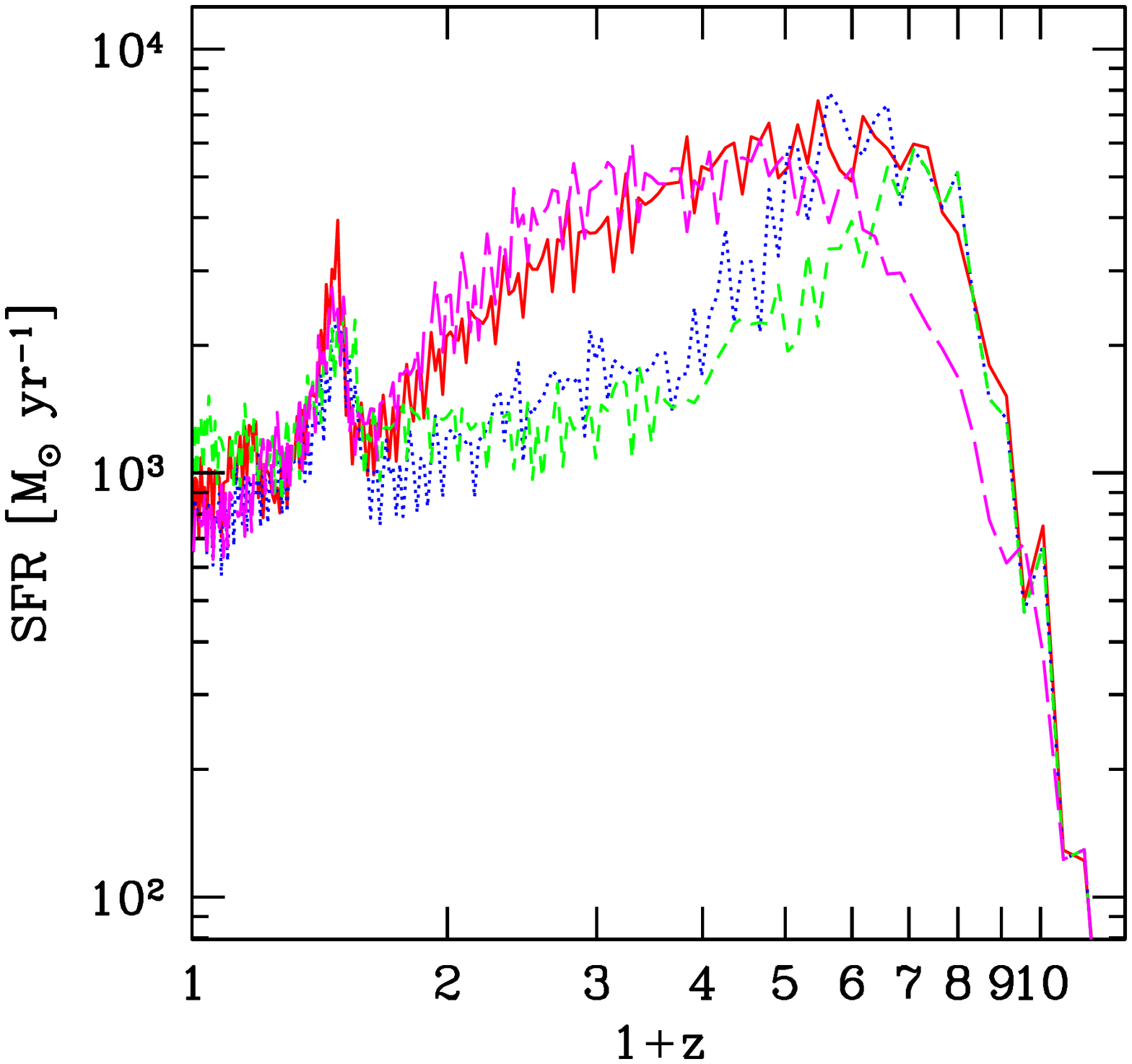}
   \epsfysize=2.5truein  \epsffile{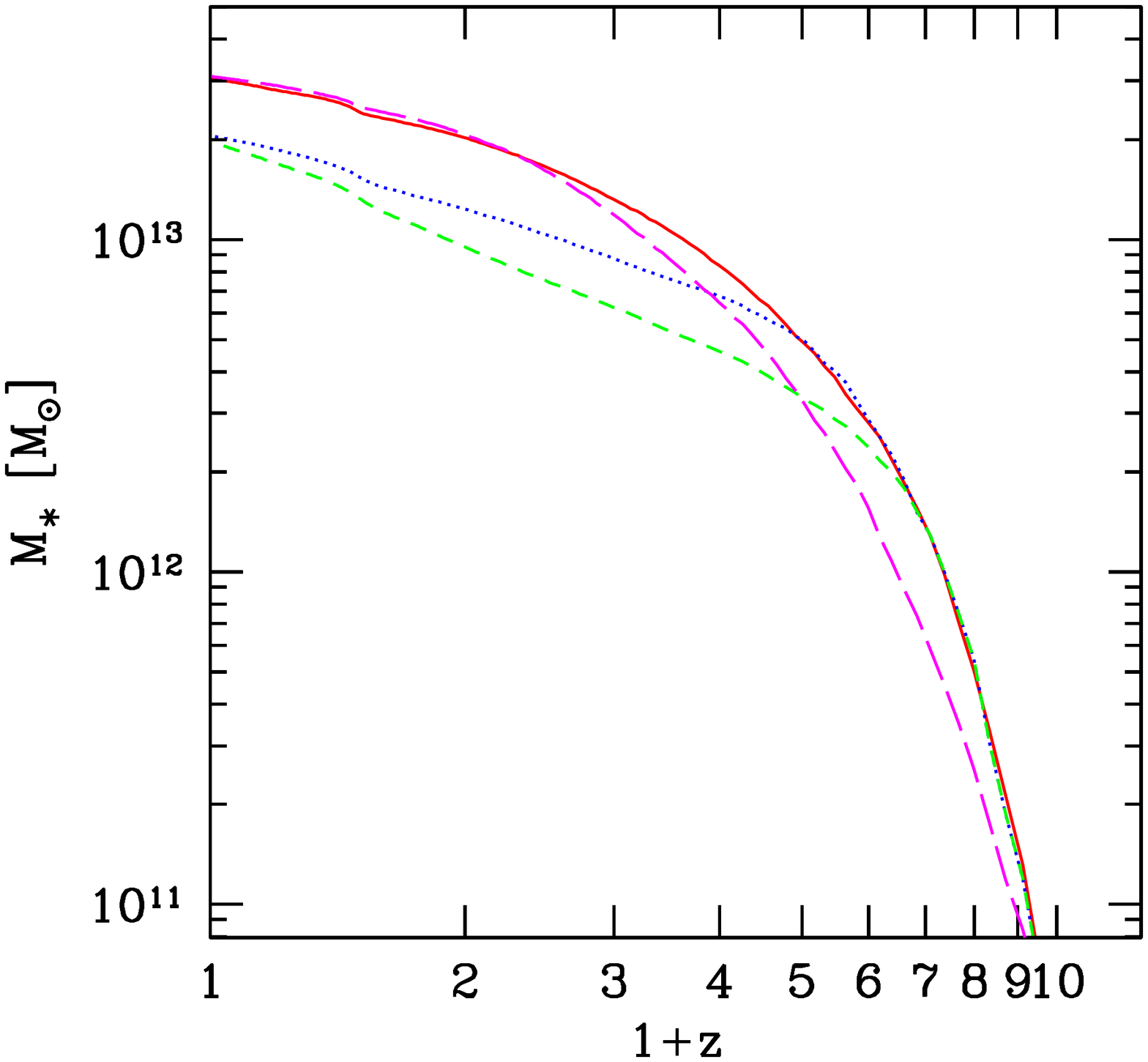}
}
\vspace{-0.6cm}
\caption{
  The star formation rate (left) and the stellar mass (right) as a
  function of redshift.  These star formation history was constructed
  using stellar particles in the lagrangian volume enclosed within the
  virial radius of the simulated cluster at z=0. Lines indicate the
  simulations that include cooling, star formation and stellar
  feedback (solid), and those with additional preheating of
  0.3keV/particle at z=4.5 (dotted), z=6 (dashed) and z=9
  (long-dashed). }
\label{profiles}
\end{figure}

\section{Discussions and Conclusions}

The overcooling problem is a serious challenge for the cluster
formation simulations. Our analysis shows that the overcooling in a
cluster as a whole is really the overcooling in its core. In other
words, the central cluster galaxies formed in simulations are too
massive and bright compared to observations.  Analysis of cluster
evolution indicates that the problem sets in early at high redshifts,
and is not limited to a single object or region. This indicates that
an additional heating mechanism is necessary for reproducing the
observed cluster properties. These results lend support to the idea
that feedback from Active Galactic Nuclei is an important missing
ingredient in the current simulations of cluster formation.




\end{document}